\documentclass[prd,twocolumn,showpacs,floatfix,superscriptaddress]{revtex4}
 \pdfoutput=1
 
\usepackage{graphicx}
\usepackage{epsfig}
\usepackage{bm}
\usepackage{amssymb}
\usepackage{float}
\usepackage{amsmath}
\usepackage{dcolumn}
\usepackage{cancel}
\usepackage[colorlinks]{hyperref}
\usepackage[usenames,dvipsnames]{color}
\hypersetup{
     breaklinks=true,
    pdfstartview={FitH},    
    colorlinks=true,       
    linkcolor=blue,          
    citecolor=red,        
    filecolor=magenta,      
    urlcolor=blue,           
    anchorcolor=green,      
    linktocpage=true
}

\providecommand{\U}[1]{\protect\rule{.1in}{.1in}}

\def\e{\mbox{\rm e}}

\newcommand{\be}{\begin{equation}}
\newcommand{\ee}{\end{equation}}

\newcommand{\mincir}{\raise
-3.truept\hbox{\rlap{\hbox{$\sim$}}\raise4.truept\hbox{$<$}\ }}
\newcommand{\magcir}{\raise
-3.truept\hbox{\rlap{\hbox{$\sim$}}\raise4.truept\hbox{$>$}\ }}

\begin{document}

\title{Unified dark sectors in scalar-torsion theories of 
gravity
}

\author{Genly Leon}
\email{genly.leon@ucn.cl}
\affiliation{Departamento de Matem\'{a}ticas, Universidad Cat\'{o}lica del 
Norte, Avda.
Angamos 0610, Casilla 1280 Antofagasta, Chile}
\affiliation{Institute of Systems Science, Durban University of Technology, PO 
Box 1334,
Durban 4000, South Africa}

\author{Andronikos Paliathanasis}
\email{anpaliat@phys.uoa.gr}
\affiliation{Institute of Systems Science, Durban University of Technology, PO 
Box 1334,
Durban 4000, South Africa}
\affiliation{Instituto de Ciencias F\'{\i}sicas y Matem\'{a}ticas, Universidad 
Austral de
Chile, Valdivia 5090000, Chile}

\author{Emmanuel N. Saridakis}
\email{msaridak@noa.gr}
\affiliation{National Observatory of Athens, Lofos Nymfon, 11852 Athens, Greece}
\affiliation{Department of Astronomy, School of Physical Sciences, University 
of 
Science and Technology of China, Hefei, Anhui 230026, China}
\affiliation{CAS Key Laboratory for Research in Galaxies and 
Cosmology, University of
Science and Technology of China, Hefei, Anhui 230026, China}

\author{Spyros Basilakos}
\email{svasil@academyofathens.gr}
\affiliation{National Observatory of Athens, Lofos Nymfon, 11852 Athens, Greece}
\affiliation{Academy of Athens, Research Center for Astronomy and Applied 
Mathematics,\\
Soranou Efesiou 4, 11527, Athens, Greece}
\affiliation{
School of Sciences, European University Cyprus, Diogenes Street, Engomi 1516 
Nicosia}

\begin{abstract}
We present a unified description of the matter and dark energy epochs, using a 
class of scalar-torsion theories. We provide a Hamiltonian description,  and by 
applying Noether’s theorem and by requiring the field equations to admit 
linear-in-momentum conservation laws we obtain two
specific classes of scalar-field potentials. We extract analytic solutions and 
we perform a detailed dynamical analysis. We show that the system possesses 
critical points that correspond to scaling solutions in which the effective, 
total equation-of-state parameter is close to zero, and points in which it is 
equal to the cosmological constant value $-1$. Therefore, during evolution, the 
Universe remains for sufficiently long times at the epoch corresponding to 
dust-matter domination, while at later times it enters the accelerated epoch and 
it eventually results in the de Sitter phase.  Finally, in contrast to other 
unified scenarios, such as Chaplygin gas-based models as well as Horndeski-based 
constructions, the present scenario is free from instabilities and pathologies 
at the perturbative level.

\end{abstract}
 
\maketitle

\section{Introduction}
\label{sec1}
According to detailed observations of different origins, the Universe has 
entered a period of accelerated expansion in the recent cosmological past. 
The simplest explanation is the cosmological constant, nevertheless, the 
corresponding problem, as well as the possibility of a dynamical nature, led to 
two main classes for its description. The first is to maintain general 
relativity and introduce the concept of dark energy, which accounts for all 
forms of new, exotic sectors that can be sources of acceleration 
\cite{Copeland:2006wr, Cai:2009zp}. The second is to attribute the new degrees 
of freedom to modifications of the gravitational interaction 
\cite{CANTATA:2021ktz, Capozziello:2011et, Nojiri:2010wj}, namely to extended 
theories that have general relativity as a limit but which in general have a 
richer structure.
Additionally, there are cumulative pieces of evidence that most of the 
universe's matter content is in the form of  Cold Dark Matter (CDM)  
\cite{Liddle:1993fq, Planck:2018vyg, Abdalla:2022yfr}. 
Although most cosmologists believe that dark matter should correspond to some 
particle beyond the Standard Model, the fact that it has not been directly 
detected in the accelerators led to the investigation of many models in which 
dark matter can have, partially or completely, gravitational origin 
\cite{Nojiri:2006gh, Famaey:2011kh, Sebastiani:2016ras, Addazi:2021xuf}. 
Modified  theories of gravity may arise by extending the Einstein-Hilbert action 
in a suitable way, such as in $F(R)$  \cite{DeFelice:2010aj} and $F(G)$ 
\cite{Nojiri:2005jg,DeFelice:2008wz} gravity, in Lovelock construction 
\cite{Lovelock:1971yv,Deruelle:1989fj}, in Horndeski gravity 
\cite{Horndeski:1974wa}, in  generalized galileon theories 
\cite{DeFelice:2010nf,Deffayet:2011gz}, etc. 
Nevertheless, one can construct gravitational modifications starting from  the 
equivalent  torsional
formulation of gravity \cite{Pereira,Maluf:2013gaa} and build theories such as 
$F(T)$ gravity
\cite{Cai:2015emx,Ferraro:2006jd,Linder:2010py}, $F(T,T_{G})$ gravity
\cite{Kofinas:2014owa}, $F(T,B)$ gravity \cite{Bahamonde:2015zma}, etc.
In this framework one can also introduce scalar fields, i.e. constructing   
scalar-torsion theories \cite{Geng:2011aj}, allowing for non-minimal  
\cite{Geng:2011aj,Geng:2011ka,Gonzalez-Espinoza:2020jss, Paliathanasis:2021nqa, 
Gonzalez-Espinoza:2021qnv,Toporensky:2021poc} or derivative  
\cite{Kofinas:2015hla} couplings with torsion, or more general constructions 
\cite{Geng:2012vtr,Skugoreva:2014ena,Jarv:2015odu,Skugoreva:2016bck,
Hohmann:2018rwf,Hohmann:2018vle,Hohmann:2018ijr,Hohmann:2018dqh,
Emtsova:2019qsl}, than can even be the teleparallel version of Horndeski 
theories 
\cite{Bahamonde:2019shr,Bahamonde:2020cfv,Bahamonde:2021dqn,Bernardo:2021izq}.
  
On the other hand, a large amount of research has been devoted to obtaining a  
unified description of the Universe, namely introducing a single component that 
can behave as dust matter at early and intermediate times, and as the 
acceleration source at late times. The typical example of such classes is the 
Chaplygin gas cosmology \cite{Bilic:2001cg,Bento:2002ps,Dev:2002qa}, in which 
one introduces by hand an exotic fluid with a peculiar equation-of-state 
parameter, that is close to zero at early times,  it is progressively 
increasing, acquiring a value around $-0.7$ today, as it is required by the 
observed total equation of   state of the Universe at present, and finally in 
the future results to $-1$, i.e to de Sitter phase. Similarly, one can apply 
Generalized Galileons/Horndeski theories 
\cite{Horndeski:1974wa,DeFelice:2010nf,Deffayet:2011gz} and require the extra 
scalar field to describe both the matter and dark-energy sectors in a unified 
way \cite{Koutsoumbas:2017fxp}.
However, although  the 
generalized Chaplygin gas can indeed describe the evolution of the Universe at 
the background level \cite{Bento:2002yx,Bento:2005un}, it may lead to 
perturbative 
instabilities  \cite{Sandvik:2002jz}, which then require the introduction of 
extra mechanisms to cure them, such as  small entropy perturbations 
\cite{Farooq:2010xm,Gorini:2007ta} or baryonic matter that can improve the 
behavior of the matter power spectrum 
\cite{Debnath:2004cd,BouhmadiLopez:2004mp,Setare:2007mp}. Similarly, 
in Horndeski-based theories of dark-sector unification, perturbative 
instabilities related to the sound-speed square may arise too 
\cite{Koutsoumbas:2017fxp,Babichev:2007dw,Deffayet:2010qz,Easson:2013bda}.

In this work, we want to present a unified description of the matter and dark 
energy epochs, using not a peculiar, exotic fluid, but a class of 
scalar-torsion theories.   Although such constructions are usually applied 
for the description of only the dark energy sector, in this work, we apply 
the Noether symmetry approach to constructing suitably models which give rise 
to 
an effective cosmic fluid whose equation-of-state behaves as the pressureless 
matter at early times, and as dark energy at late ones. As we 
will see, the unified description can indeed be obtained, and this is 
achieved without the presence of instabilities at the perturbative level.

The plan of the work is the following. In Section \ref{sec2} we briefly 
review the scalar-torsion theories of gravity and in Section \ref{sec3} we 
present their Hamiltonian description, focusing on the conservation laws. 
Then, in Section \ref{sec4} we extract analytic solutions, investigating the 
asymptotic dynamics by applying the dynamical system analysis. Finally, in 
Section \ref{con00} we discuss our results and present our conclusions.

\section{Scalar-torsion theories}
\label{sec2}

In the torsional formulation of gravity   one uses the tetrad fields 
${\mathbf{e}_A(x^\mu)}$, which form an orthonormal
basis  at a manifold point  $x^\mu$. In a coordinate basis they are expressed as
$\mathbf{e}_A=e^\mu_A\partial_\mu $, and the spacetime metric is 
\begin{equation}  \label{metricrel}
g_{\mu\nu}(x)=\eta_{AB}\, e^A_\mu (x)\, e^B_\nu (x),
\end{equation}
with    $\eta_{AB}={\rm diag} 
(-1,1,1,1)$ and where Greek  and Latin indices are used to denote coordinate and
tangent space respectively.
 Moreover,   one introduces the 
 Weitzenb\"{o}ck  connection
$\overset{\mathbf{w}}{\Gamma}^\lambda_{\nu\mu}\equiv e^\lambda_A\:
\partial_\mu
e^A_\nu$, and hence the corresponding torsion tensor reads as
\begin{equation}
\label{torsten}
{T}^\lambda_{\:\mu\nu}\equiv\overset{\mathbf{w}}{\Gamma}^\lambda_{
\nu\mu}-%
\overset{\mathbf{w}}{\Gamma}^\lambda_{\mu\nu}
=e^\lambda_A\:(\partial_\mu
e^A_\nu-\partial_\nu e^A_\mu).
\end{equation}
On can construct the torsion scalar by its contraction as
\begin{equation}
\label{torsiscal}
T\equiv\frac{1}{4}
T^{\rho \mu \nu}
T_{\rho \mu \nu}
+\frac{1}{2}T^{\rho \mu \nu }T_{\nu \mu\rho }
-T_{\rho \mu }^{\ \ \rho }T_{\
\ \ \nu }^{\nu \mu },
\end{equation}
which is then used  as the Lagrangian of teleparallel gravity.
In particular, writing the action
\begin{eqnarray}
\label{action0}
S = \frac{1}{16\pi G}\int d^4x e T,
\end{eqnarray}
with $e = \text{det}(e_{\mu}^A) = \sqrt{-g}$ and  $G$ the gravitational
constant, 
and performing variation   in terms of the tetrads leads to  
  the same equations of general relativity, and that is why the   
  theory at hand is named teleparallel 
equivalent of general relativity (TEGR).

One can add a scalar field $\phi$ to the above framework, resulting to the 
scalar-torsion theories of gravity. As it has been extensively discussed, 
although TEGR is equivalent with general relativity, $f(T)$ is different than 
$f(R)$ gravity, and similarly
scalar-torsion theories are in general different than scalar-tensor (i.e. 
scalar-curvature) gravity, due to the different structure of curvature and 
torsion tensors \cite{Cai:2015emx}. The simplest scalar-torsion  action    is
\cite{Geng:2011aj}
\begin{equation}
S=\int d^{4}xe\left[ \left(  1+\xi\phi^{2}\right)  T+\partial_\mu\phi 
\partial^\mu \phi    -2V\left(  \phi\right) \right]  , 
\label{ns.01}%
\end{equation}
with $V(\phi)$   the  potential of the scalar field, $\xi$ the coupling 
parameter, and where for simplicity we use units where $8\pi G=1$.

We consider  a spatially flat Friedmann--Lema\^{\i}tre--Robertson--Walker 
(FLRW)
background metric with line element
\begin{equation}
ds^{2}=-dt^{2}+a^{2}\left(  t\right)  \left(  dx^{2}+dy^{2}+dz^{2}\right).
\label{ns.02}%
\end{equation}
Hence, the two Friedmann equations become
\begin{align}
3H^{2}& =\rho_{\phi}, \label{ns.03}%
\\
-2\dot{H}-3H^{2}& =P_{\phi} \label{ns.04},
\end{align}
with $H=\frac{\dot{a}}{a}$   the Hubble function  
and where the effective energy density and pressure for the scalar field are 
written as \cite{Geng:2011aj}
\begin{align}
\rho_{\phi}  &  =\frac{1}{2}\dot{\phi}^{2}+V\left(  \phi\right)  -3\xi
H^{2}\phi^{2},\label{ns.05}\\
P_{\phi}  &  =\frac{1}{2}\dot{\phi}^{2}-V\left(  \phi\right)  +4\xi H\phi
\dot{\phi}+\xi\left(  2\dot{H}+3H^{2}\right)  \phi^{2}. \label{ns.06}%
\end{align}
Additionally, variation of (\ref{ns.01}) leads to   
 the Klein-Gordon equation
 \begin{equation}
\ddot{\phi}+3H\dot{\phi}+6\xi H\phi+V_{,\phi}\left(  \phi\right)  =0,
\label{ns.08}%
\end{equation}
which can be re-written equivalently as
\begin{equation}
\dot{\rho}_{\phi}+3H\left(  \rho_{\phi}+P_{\phi}\right)  =0. \label{ns.07}%
\end{equation}
Finally, note that we can introduce the equation-of-sate parameter for the 
scalar field as
  \begin{equation}
 w_{eff}\equiv\frac{P_{\phi}}{\rho_\phi}.
\end{equation}

 We close this section by presenting the 
minisuperspace description of the above theory. In particular, we start by 
re-writing the FLRW  line element adding the   lapse function, namely
\begin{equation}
ds^{2}=-N^{2}\left(  t\right)  dt^{2}+a^{2}\left(  t\right)  \left(
dx^{2}+dy^{2}+dz^{2}\right).  \label{ns.09}%
\end{equation}
Then, the field equations follow from the variation of the point-like
Lagrangian%
\begin{align}
& \mathcal{L}\left(  N,a,\dot{a},\phi,\dot{\phi}\right) \nonumber\\
& =\frac{1}{N}\left[
-3\left(  1+\xi\phi^{2}\right)  
a\dot{a}^{2}+\frac{1}{2}a^{3}\dot{\phi}^{2}\right]
-Na^{3}V\left(  \phi\right)  . \label{ns.10}%
\end{align}
Specifically, the first Friedmann (constraint) equation (\ref{ns.03}) is 
obtained from the
Euler-Lagrange equation~$\frac{\partial\mathcal{L}}{\partial N}=0$, while the
second Friedmann  equation  (\ref{ns.04}) is provided by the Euler-Lagrange 
equations
with respect to the scale factor $a$,~$\frac{d}{dt}\frac{\partial\mathcal{L}%
}{\partial\dot{a}}-\frac{\partial\mathcal{L}}{\partial a}=0$. Finally,  the 
scalar field equation arises from   
$\frac{d}{dt}\frac{\partial\mathcal{L}}{\partial\dot{\phi}%
}-\frac{\partial\mathcal{L}}{\partial\phi}=0$. As usual, in all the above 
equations  one can set $N\left(  t\right)  =1$ after the derivations.

\section{Hamiltonian description and conservation laws}
\label{sec3}

We continue by introducing a re-scaled scalar field $\psi\left(  t\right)  $ 
defined as
$\phi=\frac{1}{\sqrt{\xi}}\sinh\left(  \sqrt{\xi}\psi\right)  $. In the new
variables, the point-like Lagrangian (\ref{ns.10}) becomes
\begin{align}
& \mathcal{L}\left(  N,a,\dot{a},\psi,\dot{\psi}\right) \nonumber \\
& =\frac{1}{N}\cosh
^{2}\left(  \sqrt{\xi}\psi\right)  \left(  -3a\dot{a}^{2}+\frac{1}{2}a^{3}\dot
{\psi}^{2}\right)  -a^{3}NV\left(  \psi\right)  . \label{ns.11}%
\end{align}
Hence, the field equations become
\begin{small}
\begin{equation}
\cosh^{2}\left(  \sqrt{\xi}\psi\right)  \left(  -3H^{2}+\frac{1}{2}%
\dot{\psi}^{2}\right)  +V\left(  \psi\right)  =0, \label{ns.12}%
\end{equation}%
\begin{equation}
2\dot{H}+3H^{2}+\frac{1}{2}\dot{\psi}^{2}+4\sqrt{\xi}H\tanh\left(  \sqrt{\xi
}\psi\right)  \dot{\psi}-\frac{V\left(  \psi\right)  }{\cosh^{2}\left(
\sqrt{\xi}\psi\right)  }=0, \label{ns.13}%
\end{equation}%
\begin{equation}
\ddot{\psi}+3H\dot{\psi}+\sqrt{\xi}\tanh\left(  \sqrt{\xi}\psi\right)  \left(
\dot{\psi}^{2}+6H^{2}\right)  +\frac{V_{,\psi}\left(  \psi\right)  }{\cosh
^{2}\left(  \sqrt{\xi}\psi\right)  }=0, \label{ns.14}%
\end{equation}
\end{small}
where $H=\dot a/(a N)$, and we set $N\left(  t\right)  =1$ in the end.
Therefore, the energy density and   pressure for the scalar field are defined
as
\begin{equation}
\rho_{\psi}=\frac{1}{2}\dot{\psi}^{2}+ \frac{V\left(  \psi\right)  }%
{\cosh^{2}\left(  \sqrt{\xi}\psi\right)  }, \label{ns.15}%
\end{equation}%
\begin{equation}
P_{\psi}=\frac{1}{2}\dot{\psi}^{2}+4\sqrt{\xi}H\tanh\left(  \sqrt{\xi}%
\psi\right)  \dot{\psi}-\frac{V\left(  \psi\right)  }{\cosh^{2}\left(
\sqrt{\xi}\psi\right)  }, \label{ns.16}%
\end{equation}
while the equation-of-sate parameter  becomes
  \begin{equation}
 w_{eff}\equiv\frac{P_{\psi}}{\rho_\psi}.
 \label{weff2}
\end{equation}
Since   $\psi$ is just the re-scaled $\phi$,  the physical properties of the 
scenario are the same as the initial scalar-torsion theory.

The cosmological   equations (\ref{ns.12})-(\ref{ns.15}) form an autonomous
dynamical system described by the point-like Lagrangian (\ref{ns.11}), where 
equation
(\ref{ns.12}) can be seen as the energy conservation law  for the 
Euler-Lagrange 
equations (\ref{ns.13}), (\ref{ns.14}). For the Lagrangian function 
(\ref{ns.11}) 
we define the  generalized momenta by
$p_{i}=\frac{\partial {\mathcal {L}}}{\partial {\dot {q}}^{i}}$, where 
$q^i\in\{a, \psi\}$, $p_i\in\{p_a, p_\psi\}$,  namely
\begin{align}
p_{a}  &  \equiv -6a\frac{\cosh^{2}\left(  \sqrt{\xi}\psi\right)  }{N}\dot
{a},\label{ns.17}\\
p_{\psi}  & \equiv\frac{\cosh^{2}\left(  \sqrt{\xi}\psi\right)  }{N} 
a^3\dot{\psi}.
\label{ns.18}%
\end{align}
Hence, we can introduce the Hamiltonian function  
$\mathcal{H}=p_{a}\dot{a}+p_{\psi}\dot
{\psi}-\mathcal{L}$, which is written as
\begin{equation}
\mathcal{H}=\frac{N}{\cosh^{2}\left(  \sqrt{\xi}\psi\right)  }\left(
\frac{p_{\psi}^{2}}{2a^{3}}-\frac{p_{a}^{2}}{12a}\right)  +a^{3}NV\left(
\psi\right)  . \label{ns. 19}%
\end{equation}
Note that  $d{\mathcal H}/dt=0$, and as expected the first Friedmann 
(constraint) equation 
 (\ref{ns.12}) becomes $\mathcal{H}\left(  N,a,\psi,p_{a},p_{\psi
}\right)  \equiv0$.
Thus, we have the evolution Eqs. for $(\dot{a}, \dot{\psi})$ as given by
(\ref{ns.17}), (\ref{ns.18}). Additionally, Hamilton's equations ${\dot {p}}_{i} 
= - {\frac {\partial {\mathcal {H}}}{\partial q^{i}}}$, where $q^i\in\{a, 
\psi\}$, $p_i\in\{p_a, p_\psi\}$, lead to   
\begin{equation}
\frac{1}{N}\dot{p}_{a}=\frac{1}{\cosh^{2}\left(  \sqrt{\xi}\psi\right)
}\left(\frac{3}{2}\frac{p_{\psi}^{2}}{a^{4}}-\frac{p_{a}^{2}}{12a^{2}}\right) 
-3a^{2}V(\psi), \label{ns.20}%
\end{equation}%
\begin{equation}
\frac{1}{N}\dot{p}_{\psi}=\frac{\sqrt{\xi}\tanh\left(  \sqrt{\xi}%
\psi\right)  }{\cosh^{2}\left(  \sqrt{\xi}\psi\right)  
}\left(\frac{p_{\psi}^{2}}{a^{3}}-\frac{p_{a}^{2}}{6a} \right)  
-a^{3}V_{,\psi}(\psi). \label{ns.21}%
\end{equation}

To proceed with the derivation of analytic solutions for the
Hamiltonian system (\ref{ns.17}), (\ref{ns.18}), (\ref{ns.20}), (\ref{ns.21}),
we have to investigate the integrability properties of the system. Indeed, the
Hamiltonian function (\ref{ns. 19}) is a conservation law for the dynamical
system. Thus,  we must find the potential functions $V(\psi) $ for 
which the dynamical system admits additional conservation 
laws.

We are interested in studying the existence of conservation laws in the field 
equations that are linear in momentum. Specifically, we apply 
Noether's theorem to constraint the function $V(\psi) $, 
 for the field equations   to possess Noetherian
first integrals. This approach has been widely
applied in the literature, with many interesting results
\cite{Rosquist,Cotsakis:1998zk,Vakili:2008ea,Capozziello:2009te,Zhang:2009mm,
MohseniSadjadi:2012brg, Vakili:2011uz,Atazadeh:2011aa, Dong:2013rea, 
Christodoulakis:2014wba, Terzis:2014cra,
Dimakis:2013oza,Dimakis:2017kwx,Paliathanasis:2014zxa,Papagiannopoulos:2016dqw,
Paliathanasis:2011jq,Paliathanasis:2015aos}. Concerning teleparallel and 
torsional 
gravity, Noether's symmetry approach was applied in
\cite{Basilakos:2013rua,Paliathanasis:2014iva,Karpathopoulos:2017arc,
Capozziello:2016eaz, Bahamonde:2016grb}. Moreover, as has been discussed in
\cite{Basilakos:2011rx}, the application of Noether's conditions for the 
constraint of the
scalar field potential is a geometric selection rule since there exists a
relation between Noether symmetries and the collineations of the
minisuperspace for the field equations. In the following, we omit the 
calculations for the derivation of Noether symmetries and the approach that we 
follow can be found in detail in \cite{Tsamparlis:2018nyo}. 

In summary, we find two non-zero scalar field potentials, for which
the field equations admit linear-in-momentum conservation laws, and we 
separately present them in the following subsections.
  
\subsection{Potential $V_{A}\left(  \phi\right)=V_{0}\cosh^{-2}\left(
\sqrt{\xi}\psi\right)   $}

According to the Noether analysis, the first potential function that we extract 
is
$V_{A}\left(  
\phi\right)  =V_{0}\cosh^{-2}\left(
\sqrt{\xi}\psi\right)$. 
Thus, we find that the field equations admit the Noether
symmetries which provide the linear first integrals
\begin{eqnarray}
&&\!\!\!\!\!\!\!\!\! \!\!\!\!\!\!
X^{1}    =\frac{1}{a^{3}}
\!
\left[  a\cosh\!\left( \! \frac{\sqrt{6}}{2}\psi\!\right)
\partial_{a}-\sqrt{6}\sinh\!\left( \! \frac{\sqrt{6}}{2}\psi\!\right)
\partial_{\psi}\!\right]\!, \\
&&\!\!\!\!\!\!\!\!\! \!\!\!\!\! \! X^{2}    =-\frac{1}{a^{3}}\!\left[  
a\sinh\!\left( \! \frac{\sqrt{6}}{2}%
\psi\!\right)  \partial_{a}+\sqrt{6}\cosh\!\left( \! 
\frac{\sqrt{6}}{2}\psi\!\right)
\partial_{\psi}\!\right]\!, \\
&&\!\!\!\!\!\!\!\!\! \!\!\!\!\!X^{3}    =\partial_{\psi}~.
\end{eqnarray}
These Noether symmetries form the three-dimensional Lie algebra 
$\left\{
X^{1},X^{2},X^{3}\right\}$, with nonzero commutators $\left[  X^{1}%
,X^{3}\right]  =\sqrt{6}X^{2},~\left[  X^{2},X^{3}\right]  =-\sqrt{6}X^{1}$.
Moreover, the corresponding linear-in-momentum conservation laws are derived 
as
\begin{align}
& I_{1}\left(  a,\psi,p_{a},p_{\psi}\right)  \nonumber \\ &  
=\frac{1}{a^{3}}\left[
a\cosh\left(  \frac{\sqrt{6}}{2}\psi\right)  p_{a}-\sqrt{6}\sinh\left(
\frac{\sqrt{6}}{2}\psi\right)  p_{\psi}\right], \label{ns.22}\\
& I_{2}\left(  a,\psi,p_{a},p_{\psi}\right)  \nonumber \\ &  
=\frac{1}{a^{3}}\left[
a\sinh\left(  \frac{\sqrt{6}}{2}\psi\right)  p_{a}-\sqrt{6}\cosh\left(
\frac{\sqrt{6}}{2}\psi\right)  p_{\psi}\right], \label{ns.23}\\
 & I_{3}\left(  a,\psi,p_{a},p_{\psi}\right)    =p_{\psi}. \label{ns.24}%
\end{align}
For these functions we calculate $\left\{  I_{A},\mathcal{H}\right\}
\simeq\mathcal{H}$, which becomes $\left\{  I_{A},\mathcal{H}\right\} 
 \simeq0$, since the dynamical system is constrained 
\cite{Christodoulakis:2014wba,Terzis:2014cra,Dimakis:2013oza}.

\subsection{Potential $V_{B}\left(  \phi\right) =V_{0}
e^{-\lambda\psi}\cosh^{-2}\left(  \sqrt{\xi}\psi\right)  $}

The second potential function that we extract, in which the field equations
admit a linear first integral, is $V_{B}\left(  \phi\right)  =V_{0}
e^{-\lambda\psi}\cosh^{-2}\left(  \sqrt{\xi}\psi\right)$, which in the limit 
$\lambda=0$ recovers the  potential of the previous subsection.
The Noether
symmetries of the field equations  are 
\begin{align}
Y^{1}  &  =\frac{1}{\sqrt{6}}a^{\frac{3}{\sqrt{6}}\lambda-3}e^{-\frac{1}%
{2}\left(  \sqrt{6}-\lambda\right)  \psi}\left(  a\partial_{a}+\partial_{\psi
}\right), \\
Y^{2}  &  =-\frac{1}{\sqrt{6}}a^{-\frac{3}{\sqrt{6}}\lambda-3}e^{\frac{1}%
{2}\left(  \sqrt{6}+\lambda\right)  \psi}\left(  a\partial_{a}-\partial_{\psi
}\right)  ,~\\
Y^{3}  &  =a\lambda\partial_{a}+\partial_{\psi}~.
\end{align}
Furthermore, the nonzero commutators of the Noether symmetries $\left\{  
Y^{1},Y^{2}
,Y^{3}\right\}  $ are $\left[  Y^{1},Y^{3}\right]  =\sqrt{6}\left(
\frac{\lambda^{2}}{6}-1\right)  Y^{2}$ and $\left[  Y^{2},Y^{3}\right]
=-\sqrt{6}\left(  \frac{\lambda^{2}}{6}-1\right)  Y^{1}$. Therefore, the 
corresponding Noetherian first integrals are
\begin{align}
\Phi_{1}\left(  a,\psi,p_{a},p_{\psi}\right)   &  =\frac{1}{\sqrt{6}}%
a^{\frac{3}{\sqrt{6}}\lambda-3}e^{-\frac{1}{2}\left(  \sqrt{6}-\lambda\right)
\psi}\left(  ap_{a}+p_{\psi}\right), \label{ns.25}\\
\Phi_{2}\left(  a,\psi,p_{a},p_{\psi}\right)   &  
=-\frac{1}{\sqrt{6}}a^{-\frac{3}{\sqrt{6}}\lambda-3}e^{\frac{1}{2}\left(  
\sqrt{6}%
+\lambda\right)  \psi}\left(  ap_{a}-p_{\psi}\right), \label{ns.26}\\
\Phi_{3}\left(  a,\psi,p_{a},p_{\psi}\right)   &  =a\lambda p_{a}+6p_{\psi}~.
\label{ns.27}%
\end{align}

\section{Analytic solutions and asymptotic dynamics}
\label{sec4}

In this section, we will extract   analytic solutions for the
Liouville integrable scalar-field potentials. It is important to mention here
that these potential functions admit more conservation laws from the degrees
of   freedom of the Hamiltonian system, namely  the scalar field 
potentials are super-integrable.
As we observe,  for the above two scalar-field potentials the cosmological 
field
equations admit  three vector fields as additional Noether symmetries elements,
for each potential, which forms the same algebra. Consequently, potentials
$V_{A}\left(  \phi\right)$ and $V_{B}\left(  \phi\right)  $ admit as Noether
symmetries the same Lie algebra but in a different representation.

The solution approach that we shall follow is summarized in the following
steps \cite{Tsamparlis:2018nyo}: first, we define the normal coordinates; then, 
as a second step,
we  apply the canonical transformations  and we  write
the Hamilton-Jacobi equation; finally, from the action arising from the
Hamilton-Jacobi equation we reduce the field equations into a system of
two first-order differential equations. In general, this is considered to be
the general solution for the scenario at hand. Nevertheless, if it is feasible 
we will also provide a closed-form solution.

\subsection{Potential $V_{A}\left(  \phi\right)=V_{0}\cosh^{-2}\left(
\sqrt{\xi}\psi\right)   $}

For the scalar field potential $V_{A}\left(  \phi\right)  $, \ Noether
symmetry $X^{3}$ is already in normal coordinates. Hence, from (\ref{ns. 19})
and (\ref{ns.24}) we result to the Hamilton-Jacobi equations
\begin{equation}
\left\{  \frac{1}{2a^{3}}\left[  \frac{\partial S\left(  a,\psi\right)
}{\partial\psi}\right]  ^{2}-\frac{1}{12a}\left(  \frac{\partial S\left(
a,\psi\right)  }{\partial a}\right)  ^{2}\right\}  +a^{3}=0~, \label{ns.28}%
\end{equation}%
\begin{equation}
\left[ \frac{\partial S\left(  a,\psi\right)  }{\partial\psi}\right]
=I_{3}\text{~}. \label{ns.29}%
\end{equation}
Therefore, the corresponding action is calculated as
\begin{equation}
S\left(  a,\psi\right)  =I_{3}\psi+\int\sqrt{6I_{3}^{2}a^{-2}+12a^{4}}%
da+S_{0},
\label{ns.30}%
\end{equation}
which implies that the field equations are reduced to the following system
\begin{align}
\frac{1}{N}\dot{a}  &  =-\frac{6\sqrt{6I_{3}^{2}a^{-4}+12a^{2}}}{\cosh
^{2}\left(  \sqrt{\xi}\psi\right)  }\label{ns.31}\\
\frac{1}{N}\dot{\psi}  &  =\frac{I_{3}}{\cosh^{2}\left(  \sqrt{\xi}%
\psi\right)  }. \label{ns.32}%
\end{align}

The solution of the above equations is expressed in terms of the elliptic 
integral%
\begin{equation}
\psi\left(  a\right)  =-\frac{I_{3}}{6}\int\frac{da}{\sqrt{6I_{3}^{2}%
a^{-4}+12a^{2}}}. \label{ns.33}%
\end{equation}
For small values of $a$, we derive $\psi\left(  a\right)  \simeq\psi_{0}a^{3}%
$, and thus $\frac{1}{N}\frac{\dot{a}}{a}\simeq-6\sqrt{6I_{3}^{2}a^{-6}
+12}\left(  1-\sqrt{\xi}\psi_{0}a^{6}\right)  $, which leads to
\begin{align}
& H\left(  a\right)  =\frac{1}{N}\frac{\dot{a}}{a} \nonumber \\
& \simeq\sqrt{6I_{3}^{2}%
a^{-6}+12}\left(  1-\sqrt{\xi}\psi_{0}a^{6}\right)  \simeq\sqrt{6I_{3}%
^{2}a^{-6}}. \label{ns.34}%
\end{align}
As we deduce,   the early-time solution is approximated by that 
of a stiff fluid.
On the other hand, for large values of the scale factor we have  $\psi\left(  
a\right)
\simeq\psi_{0}\ln a,~\psi_{0}\simeq I_{3}$, which implies that
\begin{equation}
H\left(  a\right)  =\frac{1}{N}\frac{\dot{a}}{a}\simeq a^{-2\sqrt{\xi}\psi
_{0}}\text{. } \label{ns.35}%
\end{equation}
Hence, the late-time solution is that of an ideal gas.

\subsection{Potential $V_{B}\left(  \phi\right) =V_{0}
e^{-\lambda\psi}\cosh^{-2}\left(  \sqrt{\xi}\psi\right)  $}

For the potential $V_{B}\left(  \phi\right)  $ we apply the canonical
transformation $a=\frac{1}{6}\left(  \lambda\psi-u\right)  $, and the
point-like Lagrangian (\ref{ns.11}) becomes
\begin{align}
&\!\!\!\!\!\mathcal{L}\left(  N,u,\dot{u},\psi,\dot{\psi}\right)    
=\frac{1}{12N}%
\cosh^{2}\left(  \sqrt{\xi}\psi\right)  \nonumber \\
& \ \ \ \ \ \ \ \ \times \left[  \dot{u}^{2}-2\lambda\dot{u}\dot{\psi
}+\left(  \lambda^{2}-6\right)  \dot{\psi}^{2}\right] e^{  \frac{1}{2}\left(
\lambda\psi-u\right) } \nonumber\\
& \ \ \ \ \ \ \ \  +NV_{0}\e^{  -\frac{1}{2}\left(  u+\lambda\psi\right)  }
\cosh^{-2}\left(  \sqrt{\xi}\psi\right)  . \label{ns.36}%
\end{align}
Hence the Hamiltonian function (\ref{ns. 19}) reads%
\begin{eqnarray}
&&\!\!\!\!\!\!\!\!\!\!\!\!\!\!\!\!\!\! \mathcal{H=}\frac{Ne^{  
\frac{1}{2}\left(  \lambda\psi-u\right)  }}{\cosh^{2}\left(  
\sqrt{\xi}\psi\right)  }
\nonumber \\
&&   \!\!\!\!\!  \times
\left\{  \frac{1}%
{2}\left[ \left(  \lambda^{2}-6\right)  p_{u}^{2}+\lambda p_{u}p_{\psi
}+p_{\psi}^{2}\right] -V_{0}e^{-\lambda\psi}\right\}, \label{ns.37}%
\end{eqnarray}
where
\begin{align}
p_{u}  &  =\frac{1}{6\bar{N}}\left(  
 \dot{u}- \lambda\dot{\psi}\right),\label{ns.38}\\
p_{\psi}  &  =\frac{1}{6\bar{N}}\left[  \left(  \lambda^{2}-6\right)
\dot{\psi}- \lambda\dot{u}\right], \label{ns.39}%
\end{align}
with $N=\bar{N}\left[  \cosh^{2}\left(  \sqrt{\xi}\psi\right) e^{
\frac{1}{2}\left(  \lambda\psi-u\right)}  \right]$, namely
\begin{align}
\frac{1}{\bar{N}}\dot{u}  &  =\left(  6-\lambda^{2}\right)  p_{u}-\lambda
p_{\psi},\label{ns.40}\\
\frac{1}{\bar{N}}\dot{\psi}  &  =-p_{\psi}-\lambda p_{u}. \label{ns.41}%
\end{align}

Additionally, the conservation law $\Phi_{3}$ becomes $\Phi_{3}=p_{u}$. Hence, 
from the
Hamilton-Jacobi equation
\begin{eqnarray}
  &&
  \!\!\!\!\!\!\!\!\!\!\!\!\! 
  \left(  \lambda^{2}-6\right)  \left[  \frac{\partial
S\left(  u,\psi\right)  }{\partial u}\right]  ^{2}+\lambda\left[
\frac{\partial S\left(  u,\psi\right)  }{\partial u}\right]  \left[
\frac{\partial S\left(  u,\psi\right)  }{\partial\psi}\right] \nonumber\\
&& \ \  
+\left[
\frac{\partial S\left(  u,\psi\right)  }{\partial u}\right]  ^{2} 
-2V_{0}e^{-\lambda\psi}=0,
\label{ns.42}%
\end{eqnarray}
and the constraint equation
\begin{equation}
\left(  \frac{\partial S\left(  u,\psi\right)  }{\partial u}\right)  -\Phi
_{3}=0, \label{ns.43}%
\end{equation}
it follows that
\begin{small}
\begin{align}
& S\left(  u,\psi\right) \nonumber \\
& =u\Phi_{3}+\frac{1}{2}\int\left[  \sqrt{8V_{0}%
e^{-\lambda\psi}-3\Phi_{3}^{2}\left(  \lambda^{2}-8\right)  }-\lambda\Phi
_{3}\right]  d\psi+S_{0}. \label{ns.44}%
\end{align}
\end{small}
From the this action we calculate~$p_{u}=\Phi_{3}~$and $p_{\psi}=\frac{1}%
{2}\left[  \sqrt{8V_{0}e^{-\lambda\psi}-3\Phi_{3}^{2}\left(  \lambda
^{2}-8\right)  }-\lambda\Phi_{3}\right]  .~$ Thus,
we result to the reduced
system%
\begin{align}
\frac{1}{\bar{N}}\dot{u}  &  =-\frac{1}{2}\left[  \lambda\sqrt{8V_{0}%
e^{-\lambda\psi}-3\Phi_{3}^{2}\left(  \lambda^{2}-8\right)  }+\left(
\lambda^{2}-12\right)  \Phi_{3}\right]  ,\label{ns.45}\\
\frac{1}{\bar{N}}\dot{\psi}  &  =-\frac{1}{2}\left[  \sqrt{8V_{0}%
e^{-\lambda\psi}-3\Phi_{3}^{2}\left(  \lambda^{2}-8\right)  }+\lambda\Phi
_{3}\right]  . \label{ns.46}%
\end{align}

In the special case where $\Phi_{3}=0$~or $e^{-\lambda\psi}$ dominates, the
above system is simplified as%
\begin{equation}
\frac{1}{\bar{N}}\dot{u}=-\frac{1}{2}\left(  \lambda\sqrt{8V_{0}%
e^{-\lambda\psi}}\right)  ,~~\frac{1}{\bar{N}}\dot{\psi}=-\frac{1}{2}\left(
\sqrt{8V_{0}e^{-\lambda\psi}}\right)  , \label{ns.47}%
\end{equation}
which leads to   $\frac{d\psi}{du}=\frac{1}{\lambda}$ or 
$u=\lambda\psi$. However,
in that case $a\left(  t\right)  =\frac{1}{6}\left(  \lambda\psi-u\right)
=0$, and therefore such solutions describe the very early universe.
On the other hand, in the limit where $e^{-\lambda\psi}\rightarrow0$, the
reduced system (\ref{ns.45}), (\ref{ns.46}) becomes
\begin{align}
\frac{1}{\bar{N}}\dot{u}  &  =-\frac{1}{2}\left[  \lambda\sqrt{-3\Phi_{3}%
^{2}\left(  \lambda^{2}-8\right)  }+\left(  \lambda^{2}-12\right)  \Phi
_{3}\right]  =U_{1},\label{ns.48}\\
\frac{1}{\bar{N}}\dot{\psi}  &  =-\frac{1}{2}\left[  \sqrt{-3\Phi_{3}%
^{2}\left(  \lambda^{2}-8\right)  }+\lambda\Phi_{3}\right]  =U_{2}.
\label{ns.49}%
\end{align}
For this system,  real solutions exist when $\lambda^{2}-8<0$, and in this case 
we find $\frac
{1}{\bar{N}}\dot{u}=U_{1},~\frac{1}{\bar{N}}\dot{\psi}=U_{2}$, which yields
$\psi=\frac{U_{2}}{U_{1}}u.$

\subsection{Dynamical system analysis}

We continue our analysis by studying the general evolution of the
cosmological field equations, with the super-integrable scalar-field potentials 
given above. The approach that we follow is that of the 
$H$-normalization
\cite{Copeland:1997et}, which has been widely studied in the literature for 
various
cosmological models 
\cite{Lazkoz:2007mx,Fadragas:2014mra,Amendola:2006kh,Lazkoz:2006pa, 
Leon:2008de,Leon:2012mt,Basilakos:2019dof,Fadragas:2013ina,
Paliathanasis:2020sfe,Papagiannopoulos:2022ohv}.

\begin{table*}[ht] \centering
\setlength{\tabcolsep}{5pt}
\begin{tabular}{cllll}\hline\hline
\textbf{Point} & $\left(  x,\kappa\right)  $ & \textbf{Existence} & 
\textbf{Stability}&
\textbf{Solution}\\\hline
$P_1(\kappa^\star)$ & $\left(  0,\kappa_{1}\right)  $ &  
$\kappa_1=\kappa_1^\star$  &  
stable for 
$-\frac{1}{\sqrt{2}}<{\kappa_1^\star} <\frac{1}{\sqrt{2}}$  and &  
$w_{eff}\left(  
P_1(\kappa^\star)\right)  =-1$. de Sitter.\\
&&&  $\frac{1-{\kappa_1^\star} ^2}{\left(1-2 {\kappa_1^\star} ^2\right)^2}<\xi 
\leq 
f({\kappa_1^\star})$ 
    &\\
&&&   &\\
$P_{2}^{\left(  \pm,\pm\right)  }$ & $\left(  \pm1,\pm1\right)  $ & Always & 
$P_2^{(+,+)}$ stable for $\lambda_0>0$. &
$w_{eff}\left(  P_{2}^{\left(\varepsilon, \eta\right)  }\right)  =1+4 
\varepsilon  \eta \sqrt{\frac{2\xi}{3}}$. Scaling. \\
  &   &   & $P_2^{(+,-)}$ saddle. & \\
    &   &   & $P_2^{(-,+)}$ saddle. & \\
        &   &   & $P_2^{(-,-)}$ stable for $\lambda_0<0$. & \\
$P_{3}^{+}$ & $\left(  x, 1\right)  $ & $\lambda_0=- 2$ & Numerical elaboration 
(see text and figures)
& 
$w_{eff}\left(P_{3}^{+}\right)  =-1+2x^{2}+4\sqrt{\frac{2\xi}{3}}x.$\\
&&&& de
Sitter for $x\in \left\{0, -2 
\sqrt{\frac{2}{3}} \sqrt{\xi }\right\}$, \\
&&&& matter dominated for $x=-\frac{2\sqrt{\xi}\pm
\sqrt{3+4\xi}}{\sqrt{6}}$. \\
$P_{3}^{-}$ & $\left(  x,- 1\right)  $ & $\lambda_0= 2$ & Numerical elaboration 
(see text and figures) 
& 
 $w_{eff}\left(P_{3}^{-}\right)  =-1+2x^{2}-4\sqrt{\frac{2\xi}{3}}x.$\\
&&&&de
Sitter for  $x\in \left\{0, 2 
\sqrt{\frac{2}{3}} \sqrt{\xi }\right\}$,\\
&&&& matter dominated for  $x=\frac{2\sqrt{\xi}\pm
\sqrt{3+4\xi}}{\sqrt{6}}$. 
\\\hline\hline
\end{tabular}
\caption{Critical points and curves of critical points, their existence and 
stability conditions, and their physical features.  ${\kappa_1^\star}$ 
is the solution of the equation   $ 
2\sqrt{6\xi}\kappa_{1}^\star\sqrt{1-(\kappa_{1}^\star)^{2}}-\sqrt{6}\mu\left(  
\kappa_{1}^\star\right)   
=0$, while $f({\kappa_1^\star})=
   \frac{1}{8} \left[\frac{7-10 {\kappa_1^\star} ^2}{\left(1-2 {\kappa_1^\star} 
^2\right)^2}+2 
\sqrt{2} \sqrt{\frac{8 {\kappa_1^\star}^4-13 {\kappa_1^\star}
   ^2+5}{\left(1-2 {\kappa_1^\star} ^2\right)^4}}\right]$.}    
\label{tab1}
\end{table*}

We consider the new dimensionless variables
\begin{equation}
x=\frac{\dot{\psi}}{\sqrt{6}H},~y=\frac{V\left(  \psi\right)  }{\sqrt{3}%
\cosh\left(  \sqrt{\xi}\psi\right)  H},~\kappa=\tanh\left(  \sqrt{\xi}%
\psi\right), \label{ns.50}%
\end{equation}
and we introduce the new independent variable $\omega$, defined through $d\omega
=\frac{H}{\sqrt{1-\kappa^{2}}}dt$.   
Thus, the field equations can be transformed to  the following
algebraic-differential system:
\begin{align}
& \frac{dx}{d\omega}=\frac{1}{2}\left(  x^{2}-1\right)  \left[  2\left(
3x+\sqrt{6\xi}\kappa\right)  \sqrt{1-\kappa^{2}}-\sqrt{6}\mu\left(
\psi\right)  \right], \label{ns.51}%
\\
& \frac{d\kappa}{d\omega}=\sqrt{6\xi}x\left(  1-\kappa^{2}\right)  ^{\frac{3}%
{2}},~\label{ns.52}%
\end{align}
with $\mu\left(  \psi\right)  \equiv-\left(  \ln V\left(  \psi\right)  \right)
_{,\psi}$,
while from the constraint equation (\ref{ns.12}) it follows that
\begin{equation}
1-x^{2}-y^{2}=0\label{ns.53}.
\end{equation}

For the super-integrable scalar  potential $V_{B}\left(  \psi\right)  $,
we calculate $\mu\left(  \psi\left(  \kappa\right)  \right)  =\lambda_0
+2\tanh\left(  \sqrt{\xi}\psi\right)  $, that is $\mu\left(  \kappa\right)
=\lambda_0+2\kappa,$ where $\lambda_0$ is a constant. Hence,  for 
$\lambda_0=0$  potential
$V_{B}\left(  \psi\right)  $ reduces to the form of $V_{A}\left(  \psi\right)
$. Furthermore, the effective equation-of-state parameter for the  cosmological
fluid (\ref{weff2}) is expressed in terms of the new variables as%
\begin{equation}
w_{eff}\left(  x,\kappa\right)  =-1+2x^{2}+4\sqrt{\frac{2\xi}{3}}x\kappa.
\label{ns.54}%
\end{equation}

In Table  \ref{tab1} we summarize the results of the dynamical analysis, namely 
the critical points and curves of critical points, their existence and 
stability conditions, and their physical features quantified by the  
equation-of-state parameter of the total  cosmological
fluid.

The set of critical points $P_1(\kappa^\star)$, where   ${\kappa_1^\star}$ 
is the solution of the equation   $ 
2\sqrt{6\xi}\kappa_{1}^\star\sqrt{1-(\kappa_{1}^\star)^{2}}-\sqrt{6}\mu\left(  
\kappa_{1}^\star\right)   
=0$,
  describe de Sitter universes, in which $w_{eff}\left(  
P_1(\kappa^\star)\right)  =-1$. 
The corresponding
eigenvalues of the linearized system are derived as
\begin{small}
\[
e^{\pm}=\frac{\sqrt{1-{\kappa^\star}^2}}{2}\! \left\{\!-3 \pm \sqrt{3\!\left[8 
\left(2 {\kappa^\star} 
^2\!-\!1\right) \xi +8 \sqrt{\left(1\!-\!{\kappa^\star}^2\right) \xi 
}+3\right]\!}\right\},
\]
\end{small}
and thus  points $P_1(\kappa^\star)$ are stable, namely late-time attractors, 
for 
$-\frac{1}{\sqrt{2}}<{\kappa^\star} <\frac{1}{\sqrt{2}}, \frac{1-{\kappa^\star} 
^2}{\left(1-2 
{\kappa^\star} ^2\right)^2}<\xi \leq
   \frac{1}{8} \left[\frac{7-10 {\kappa^\star} ^2}{\left(1-2 {\kappa^\star} 
^2\right)^2}+2 
\sqrt{2} \sqrt{\frac{8 {\kappa^\star} ^4-13 {\kappa^\star}
   ^2+5}{\left(1-2 {\kappa^\star} ^2\right)^4}}\right]$.

Additionally, points $P_{2}^{\left(\varepsilon, \eta \right)  }$, with
$\varepsilon=\pm 1, \eta=\pm1$, describe scaling solutions 
with
$w_{eff}\left(  P_{2}^{\left( \varepsilon, \eta \right)  }\right)  =1+4 
\varepsilon  \eta \sqrt{\frac{2\xi}{3}
}$.

\begin{figure*}[ht]
    \centering
    \includegraphics[scale=0.65]{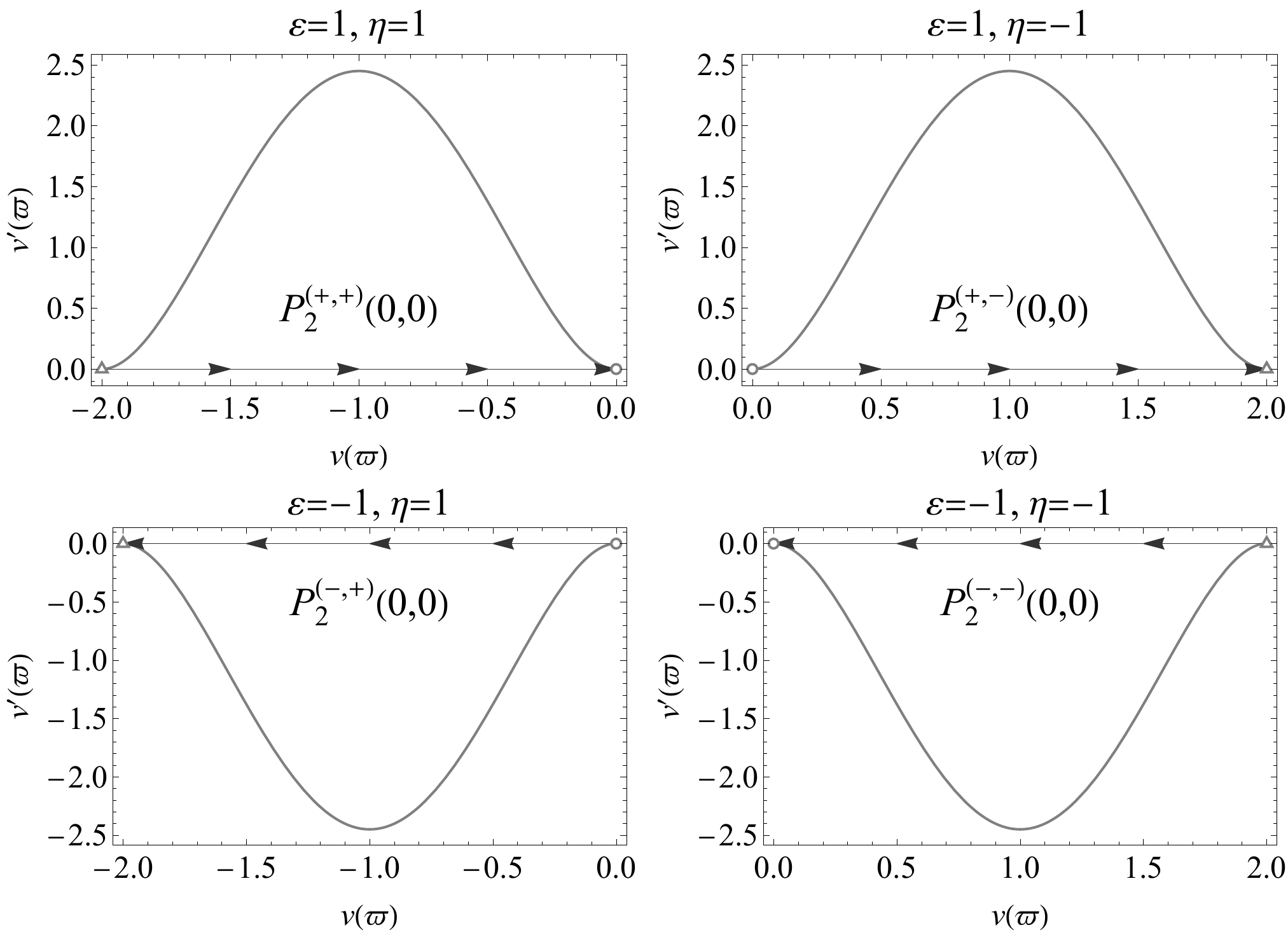}
    \caption{{\it{One-dimensional flow of the dynamical system \eqref{eq75} 
for   $\varepsilon=\pm1, \eta=\pm1$. The points of interest $P_2^{(\pm, 
\pm)}$ are represented by open circles. Note that the center manifolds  of 
the points $P_2^{(+,+)}$ and $P_2^{(-,-)}$ are stable, whereas  the center 
manifolds 
 of the points $P_2^{(+,-)}$ and $P_2^{(-,+)}$ are unstable.  }}}
    \label{fig:1D}
\end{figure*}

\begin{figure*}[ht]
\centering\includegraphics[width=0.6\textwidth]{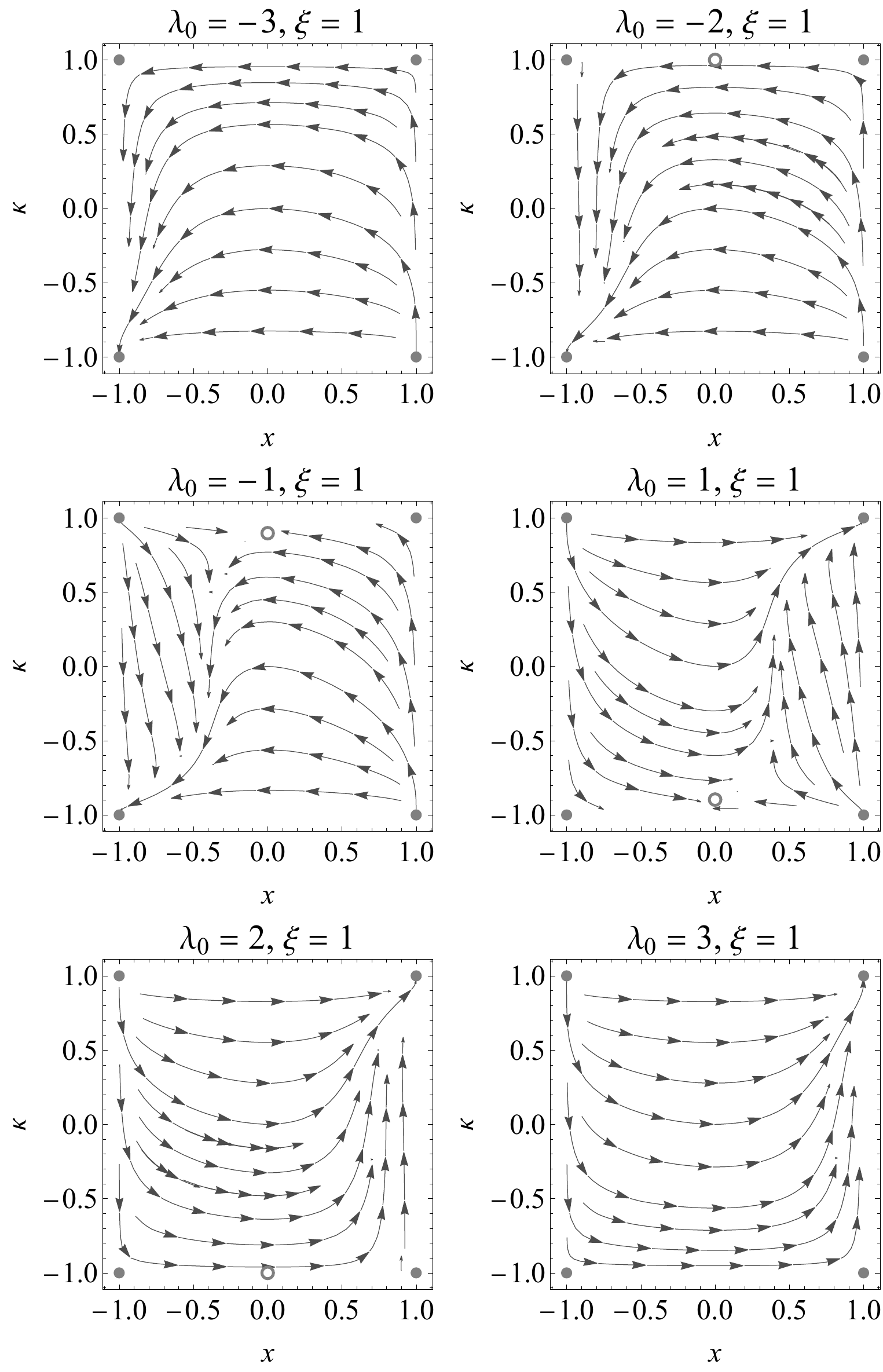}
\caption{{\it{Phase-space
portraits for the dynamical     system (\ref{ns.51}), (\ref{ns.52}),
for the value $\xi=1$ and various
values of the free parameter $\lambda_{0}$. Closed circles denote the 
equilibrium points $P_2^{\left(\pm 1, \pm 1\right)}$ and open circles denote 
the 
equilibrium point $P_1(\kappa^\star)$. For
$\lambda_{0}<0$, point $P_{2}^{\left(-,-\right)  }$ is the stable, late-time 
attractor, while
for $\lambda_{0}>0$, point $P_{2}^{\left(+,+\right)  }$ is the stable 
late-time solution of the system, and $P_1(\kappa^\star)$ is a saddle. }}}
\label{fig1}%
\end{figure*}
The linearized matrix for the vector field \eqref{ns.51}-\eqref{ns.52} is 
\begin{equation}
  J=  \left[
\begin{array}{cc}
J_{11} & J_{12}\\
J_{2 1}& J_{2 2} \\
\end{array}
\right],
\end{equation}
where
\begin{eqnarray}
&&
\!\!\!\!\!\!
J_{1 1}=  -3 \sqrt{1-\kappa ^2}+9 \sqrt{1-\kappa ^2} 
x^2\nonumber\\
&&  \ \ \ \ \ +\sqrt{6} x \left[2 \kappa  \left(\sqrt{\xi -\kappa ^2 
\xi
   }-1\right)-\lambda_0 \right]\nonumber\\
   &&\!\!\! \!\!\!J_{1 2 }=  
-\frac{\left(x^2-1\right) \left\{\sqrt{6} \left[\left(2 \kappa ^2-1\right) 
\sqrt{\xi
   }+\sqrt{1-\kappa ^2}\right]+3 \kappa  x\right\}}{\sqrt{1-\kappa 
^2}}\nonumber\\
&&\!\!\!\!\!\!J_{2 1}= \sqrt{6} \left(1-\kappa ^2\right)^{3/2} \sqrt{\xi }  
\nonumber\\
   &&\!\!\!\!\!\!J_{2 2}=-3 \sqrt{6} \kappa  x \sqrt{\xi -\kappa ^2 \xi }.
   \end{eqnarray}
The 
Hartman-Grobman theorem or Center Manifold theorem relies on the fact that a 
system evolving in time as $u(t)\in \mathbb {R} ^{n}$ must satisfy the 
differential equation $du/dt=f(u)$ for some smooth map $f:\mathbb {R} ^{n}\to 
\mathbb {R} ^{n}$ \cite{normally,Strogatz}. This is clearly not the case for 
the 
vector field \eqref{ns.51}-\eqref{ns.52} at $\kappa=\pm1$, since $J$ is not 
continuous at $\kappa=\pm 1$ due to the fact that  $J_{1 2}$ is not bounded at 
$\kappa=\pm 1$. 
Therefore, in order to conclude on the stability of  
$P_2^{\left(\varepsilon, \eta\right)}$ we need to perform a numerical 
elaboration. Introducing the variables  
 \begin{align}
 u= x - \varepsilon,  \ \ \   v= \kappa - \eta,
 \end{align}
 we
result to the dynamical system
\begin{align}
\frac{du}{d\omega}=  & 
   -\sqrt{\frac{3}{2}} \lambda_0  u (u+2 \varepsilon) \nonumber \\
 &  + 3 u \left(u^2+3 u \varepsilon +2\right) \sqrt{-v (2 \eta +v)} \nonumber 
\\ 
 &  +\sqrt{6} u (u+2 \varepsilon ) (\eta +v)
   \left[\sqrt{-\xi  v (2 \eta +v)}-1\right], \label{Newns.51}%
\\
 \frac{dv}{d\omega} = & \frac{\sqrt{6} (u+\varepsilon
   ) [-\xi v (2 \eta +v)]^{3/2}}{\xi}, \label{Newns.52}%
\end{align} 
where we assume the $v$-range: $0\leq v\leq 2$ for $\eta =-1$, or $-2\leq v\leq 
0$ for $\eta =+1$, as the physical ones. Although the corresponding 
linearization 
matrix is not bounded and is not continuous at $(u,v)=(0,0)$, we can obtain 
partial information about the stability at the origin by studying the invariant 
set $u=0$.  The dynamics at  the invariant set $u=0$ is given by 
\begin{equation}
   \frac{d v}{d \omega}= \frac{\sqrt{6} \varepsilon  [-\xi  v (2 \eta
   +v)]^{3/2}}{\xi }.
\end{equation}
By assuming $\xi>0$, and re-scaling time by $\frac{d}{d\varpi} = \sqrt{-v (2 
\eta +v)/\xi } \frac{d}{d \omega}$, 
we obtain 
\begin{equation}
     \frac{d v}{d\varpi}= \sqrt{6} \varepsilon  v^2 (2 \eta
   +v)^2, \label{eq75}
\end{equation}
where $0\leq v\leq 2$ for $\eta =-1$, or $-2\leq v\leq 0$ for $\eta =+1$, is 
the 
physical domain for $v$. 
In a one-dimensional phase space, $ \frac{d v}{d\varpi}>0$ implies that the 
arrow 
is directed to the right, and $ \frac{d v}{d\varpi}<0$ implies that the arrow 
is 
directed to the left \cite{Strogatz}. Hence, from the analysis of the 
corresponding one-dimensional flow it follows that the center manifolds  of the 
points $P^{(+,+)}$ and $P^{(-,-)}$ are stable, whereas  the center manifolds  
of the points $P^{(+,-)}$ and $P^{(-,+)}$ are unstable. 
These results are illustrated in Fig. \ref{fig:1D}.

\begin{figure*}[ht]
\centering\includegraphics[width=0.6\textwidth]{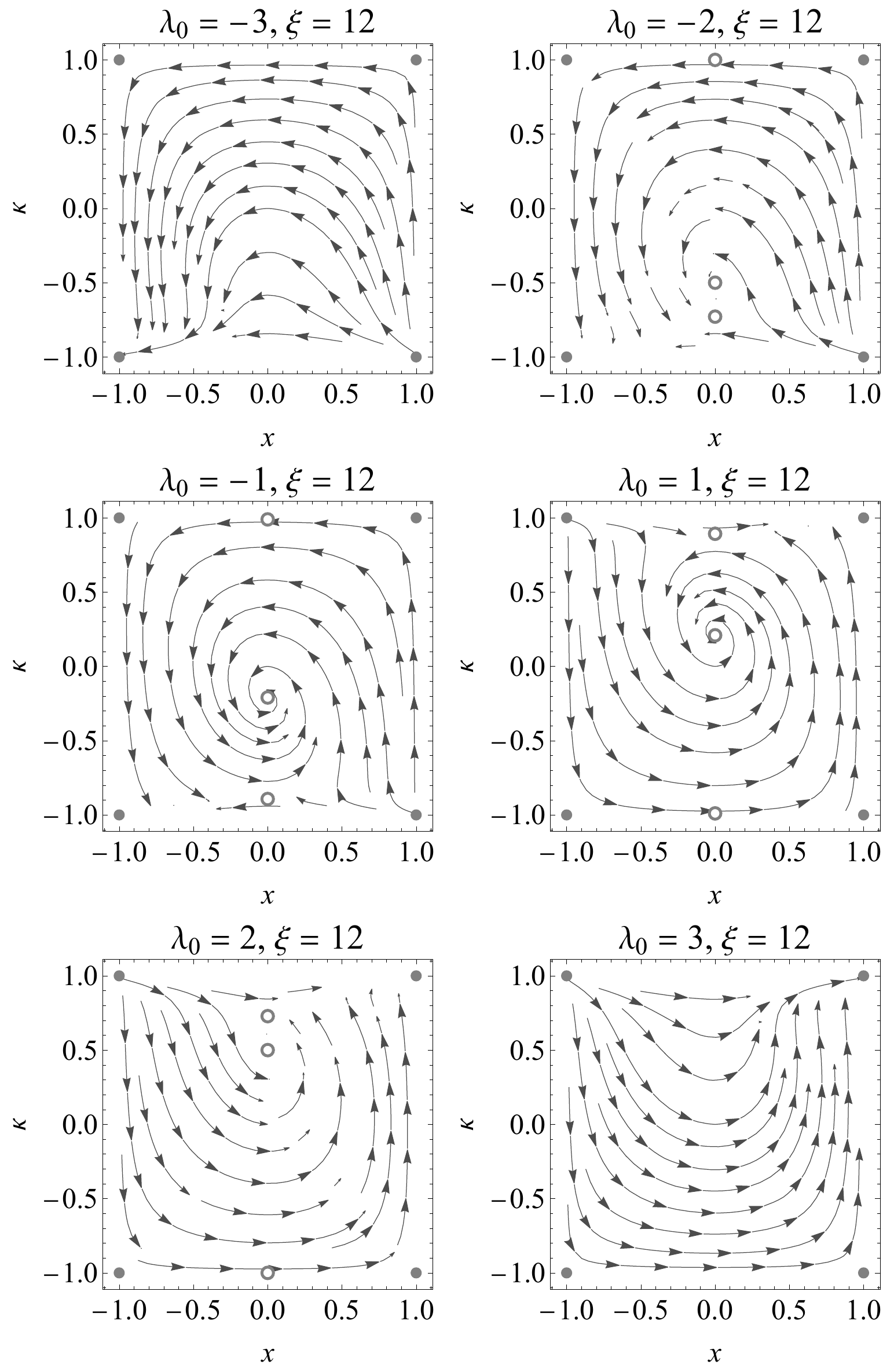}
\caption{{\it{Phase-space
portraits for the dynamical     system (\ref{ns.51}), (\ref{ns.52}),
for the value $\xi=12$ and various
values of the free parameter $\lambda_{0}$. Closed circles denote the 
equilibrium points $P_2^{\left(\pm 1, \pm 1\right)}$ and open circles denote the 
equilibrium points of type $P_1(\kappa^\star)$. For
$\lambda_{0}<0$, point $P_{2}^{\left(-,-\right)  }$ is the stable, late-time 
attractor, while
for $\lambda_{0}>0$, point $P_{2}^{\left(+,+\right)  }$ is the stable 
late-time solution of the system. One of the equilibrium points of type 
$P_1(\kappa^\star)$  is stable (open circle in the middle of some graphs). }}}
\label{fig2}%
\end{figure*}
We continue by 
presenting the phase-space
diagram for the dynamical system (\ref{ns.51}), (\ref{ns.52}) for various
values of the free parameters $\xi~$and $\lambda_{0}$ 
in Figs. \ref{fig1} and \ref{fig2}. As we observe in both figures  for
$\lambda_{0}>0$, point $P_{2}^{\left(  +,+\right)  }$ is the attractor, while
for $\lambda_{0}<0$, point $P_{2}^{\left(  -,-\right)  }$ is the stable 
late-time solution of the system. For $\xi=1$, there is an equilibrium point 
$P_1(\kappa^\star)$, which is   saddle (see Fig.  \ref{fig1}). Furthermore, for 
 $\xi=12$, there are three equilibrium points of type $P_1(\kappa^\star)$, 
and one of them can be stable (open circle in the middle  of some graphs in 
Fig. \ref{fig2}).

Lastly, let us examine the special case where $\lambda=-2$. In this case for 
the 
additional curve of critical points $P_{3}^+$ we find
that $w_{eff}\left(P_3^+\right)  =-1+2x^{2}+4\sqrt{\frac{2\xi}{3}}x$.
These points describe a de Sitter solution for $x\in \left\{0, -2 
\sqrt{\frac{2}{3}} \sqrt{\xi }\right\}$. Hence,  the points of $P_{3}^+$  
describe 
matter-dominated solutions when $x^{\pm}=-\frac{2\sqrt{\xi}\pm
\sqrt{3+4\xi}}{\sqrt{6}}$. Moreover, as far as $\xi$ is concerned, for $x^{+}$ 
it follows $0<\xi<\frac{3}{32}$, while for $x^{-}$ we obtain $0<\xi$.  
Similarly, for the additional curve of points $P_{3}^-$ we find
that $w_{eff}\left(P_3^{-}\right)  =-1+2x^{2}-4\sqrt{\frac{2\xi}{3}}x$.
These points describe de Sitter solutions for $x\in \left\{0, 2 
\sqrt{\frac{2}{3}} \sqrt{\xi }\right\}$. $P_{3}^-$   describe  matter-dominated 
universes when $x^{\pm}=\frac{2 \sqrt{\xi }+\sqrt{4 \xi
   +3}}{\sqrt{6}}$, and    for $x^{+}$ 
it follows $0<\xi<\frac{3}{32}$, while for $x^{-}$ we acquire $0<\xi$.

\subsection{Unified dark sectors}

We have now all the information to proceed to a unified description of 
the dark sectors. In particular, as we saw, the scenario of the scalar-torsion 
theory at hand possesses critical points in which the total, effective 
equation-of-state parameter of the universe behaves as dust matter and 
critical points in which it behaves as dark energy. Thus, with the single 
sector, we can describe both the matter and late-time acceleration epochs.

\begin{figure}[t]
\centering\includegraphics[width=0.45\textwidth]{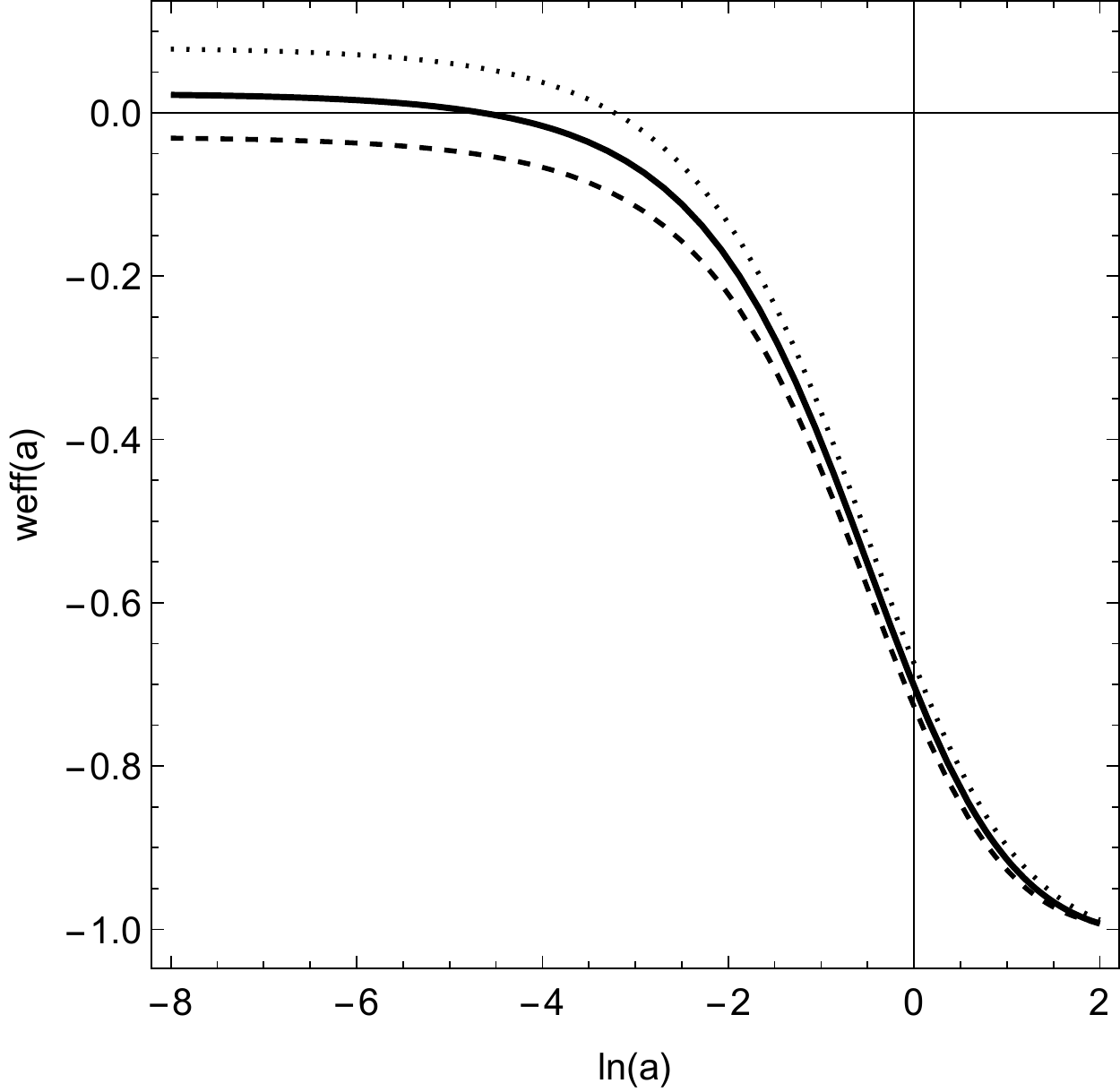}
\caption{{\it{Evolution 
of  effective, total equation-of-state parameter $w_{eff}$ of the cosmological 
fluid in scalar-torsion theory, for $\lambda=2$ and $\xi=0.08$
(solid line), $\ \xi=0.09$ (dotted line) and $\xi=0.1$ (dashed line). We have 
imposed $w_{eff}\approx-0.7$ at present time $a=a_0=1$ as required by 
observations.}}}
\label{wwfig}%
\end{figure}

In Fig. \ref{wwfig} we depict the  evolution of the effective 
equation-of-state parameter $w_{eff}$  for various choices of  $\lambda$ and 
$\xi$. As we can see, the universe at intermediate times remains around the 
scaling solution in which $w_{eff}$  is close to zero, and hence   
this era corresponds to the dust matter-dominated phase. As time passes the 
universe enters into an accelerated phase in which $w_{eff}$ becomes smaller 
than $-1/3$, and at present time it becomes equal to $-0.7$ as required by 
observations. Finally, at asymptotically late times,  the Universe will 
result in a de Sitter phase.
Hence, using models of scalar-torsion theory we succeeded in describing the 
matter and late-time accelerated eras with a single sector, which was the goal 
of the present work.

\section{Conclusions}
\label{con00}

We presented  a unified description of the matter and dark 
energy epochs, using not a peculiar, exotic fluid, but a class of 
scalar-torsion theories. In particular, we started from the subclass  
of such theories in which a scalar field is non-minimally coupled with the 
torsion scalar and we provided a Hamiltonian description, focusing on the 
conservation laws. Then, by applying Noether’s theorem and by requiring the 
 field equations to admit linear-in-
momentum conservation laws we extracted two classes of potentials for the 
scalar field.

For the two scalar potentials we extracted analytic solutions and we performed 
a detailed dynamical analysis to extract the critical points and their 
properties and thus the global feature of the Universe evolution independently 
of the initial conditions. As we saw, the system possesses critical points that 
correspond to scaling solutions in which the effective, total  equation-of-state 
parameter is close to zero, and points in which it is equal to the cosmological 
constant value $-1$. Therefore, during its evolution, the Universe remains for 
sufficiently long times around the scaling solutions, i.e. in the epoch 
corresponding to dust-matter domination, while at later times  $w_{eff}$ 
decreases and becomes smaller 
than $-1/3$, which marks the onset of the acceleration. Then, at present 
it is equal to $-0.7$, as required by observations, while at asymptotically late 
times the Universe results in the de Sitter phase. In summary, using  
scalar-torsion theory we succeeded in describing the matter and late-time 
accelerated eras with a single sector.

We close this section by referring to the additional important advantage of the 
scenario at hand, which is related to the stability at the perturbation level. 
As we mentioned in the Introduction, although a unification of the dark sectors 
can be obtained through Chaplygin gas-based models as well as in 
Horndeski-based constructions, in both cases perturbative instabilities, and 
pathologies related to the sound-speed square, may appear at the perturbation 
level. On the contrary, the scalar-tensor theories applied in this work is known 
to be free from instabilities and pathologies at the 
perturbative level 
\cite{Hohmann:2017qje,DAgostino:2018ngy,Gonzalez-Espinoza:2021mwr,
Bahamonde:2021dqn,Toporensky:2021poc}. This feature acts   in favor of the 
robustness of the present scenarios. We should further confront the 
perturbations of the   scenario at hand with growth data too, however, this 
important investigation lies beyond the scope of the present work and it is 
left for a future project.

\begin{acknowledgments}
 GL was funded by Vicerrector\'{\i}a de
Investigaci\'{o}n y Desarrollo Tecnol\'{o}gico at Universidad Catolica del
Norte. In addition, the research of AP was supported in part by the National
Research Foundation of South Africa (Grant Numbers 131604). The authors 
acknowledge participation in the COST Association Action CA18108 ``Quantum 
Gravity Phenomenology in the Multimessenger Approach''.
\end{acknowledgments}

\end{document}